\documentclass[letterpaper,twocolumn,prl,aps,showpacs,groupedaddress,floatfix]{revtex4-1}
\usepackage[latin1]{inputenc}
\usepackage{bm}
\usepackage[usenames]{color}
\usepackage{multirow}
\usepackage{amssymb}
\usepackage{amsbsy}
\usepackage{amsmath}
\usepackage{stmaryrd}
\usepackage{graphicx}
\usepackage{epsfig}
\usepackage{placeins}
\usepackage{ulem}
\usepackage{bbold}
\usepackage{braket}
\usepackage{blindtext}
\usepackage[colorlinks,linkcolor=blue,citecolor=blue,urlcolor=blue]{hyperref}
\makeatletter

\begin{document}
\title{Impact of eigenstate thermalization on the route to equilibrium}

\author{Jonas Richter}
\email{jonasrichter@uos.de}
\affiliation{Department of Physics, University of Osnabr\"uck, D-49069 
Osnabr\"uck, Germany}

\author{Jochen Gemmer}
\email{jgemmer@uos.de}
\affiliation{Department of Physics, University of Osnabr\"uck, D-49069 
Osnabr\"uck, Germany}

\author{Robin Steinigeweg}
\email{rsteinig@uos.de}
\affiliation{Department of Physics, University of Osnabr\"uck, D-49069 
Osnabr\"uck, Germany}

\date{\today}

\begin{abstract}
The eigenstate thermalization hypothesis (ETH) and the theory of linear 
response (LRT) are celebrated cornerstones of our understanding of the physics 
of many-body quantum systems out of equilibrium. While the ETH provides 
a generic mechanism of thermalization for states arbitrarily far from 
equilibrium, LRT extends the successful concepts of statistical mechanics to 
situations close to equilibrium. In our work, we connect these cornerstones to 
shed light on the route to equilibrium for a class of properly prepared states. 
We unveil that, if the off-diagonal part of the ETH applies, then the relaxation 
process can become independent of whether or not a state is close to 
equilibrium. Moreover, in this case, the dynamics is generated by a single 
correlation function, i.e., the relaxation function in the context of LRT. Our 
analytical arguments are illustrated by numerical results for idealized models 
of random-matrix type and more realistic models of interacting spins on a 
lattice. Remarkably, our arguments also apply to integrable quantum systems 
where the diagonal part of the ETH may break down.
\end{abstract}

\maketitle

{\it Introduction.} Both, equilibration and thermalization are omnipresent 
phenomena in nature. Simple examples in everyday life are a cup of hot coffee 
which cools down to room temperature, or an inkblot in water which spreads in 
the entire liquid. Even though the irreversible route to equilibrium 
occurs in any macroscopic and ordinary situation, the underlying microscopic 
laws of physics are reversible. In fact, the emergence of phenomenological
relaxation from truly microscopic principles such as the Schr\"odinger equation 
is not satisfactorily understood up to date. While this fundamental question 
has a long and fertile history, it has been under intense scrutiny in the last 
decade \cite{polkovnikov2011, eisert2015, nandkishore2015, gogolin2016, 
dalessio2016}. This upsurge of interest is also related to the advent of 
experiments on cold atomic gases \cite{bloch2005, langen2015}, the development 
of sophisticated numerical methods \cite{schollwoeck2011}, as well as the 
introduction of fresh theoretical concepts such as quantum typicality of pure 
states \cite{popescu2006, goldstein2006, reimann2007} and the eigenstate 
thermalization hypothesis (ETH) \cite{deutsch1991, srednicki1994, rigol2005}.
In particular, the ETH has become a cornerstone of our understanding of the 
mere existence of thermalization, but much less is known on the route to 
equilibrium as such~\cite{khatami2013, reimann2016, garciapintos2017}.

An obvious problem in this context is the absence of an universal approach to 
the time evolution of quantum many-body systems out of equilibrium. Of course, 
close to equilibrium, a powerful strategy is provided by the theory of linear 
response (LRT) \cite{kubo1991}. Further away from equilibrium, however, this 
theory is naturally expected to break down and the dynamics might drastically 
change for states far from equilibrium. In this respect, our work reports an 
unexpected and intriguing picture. Following earlier ideas developed by 
Srednicki \cite{srednicki1999}, we establish a link between LRT and ETH 
for specific non-equilibrium setups introduced below in detail. We unveil that, 
if the off-diagonal part of the ETH holds, then the relaxation process can 
become independent of whether or not a state is close to equilibrium. Moreover, 
in this case, the time evolution is generated by a single correlation function, 
i.e., the relaxation function in the context of LRT. Our analytical arguments 
are also confirmed numerically for two different models.
\begin{figure}[tb]
\centering
\includegraphics[width=0.95\columnwidth]{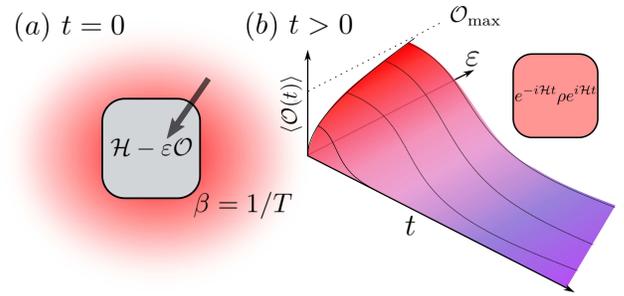}
\caption{(Color online) Sketch of the setup. (a) A static force of strength 
$\varepsilon$ gives rise to an additional potential, described by an operator 
$\cal O$. Because of the presence of a heat bath at inverse temperature $\beta = 
1/T$, thermalization to the density matrix $\rho = \exp[-\beta( {\cal H} - 
\varepsilon {\cal O})]/ {\cal Z}$ occurs. (b) Static force and heat bath are 
both removed and $\rho$ undergoes unitary time evolution with respect to 
$\mathcal{H}$. We discuss the dependence of the relaxation curve $\langle {\cal 
O}(t) \rangle$ on the perturbation $\varepsilon$ outside the regime of small 
$\varepsilon$.} \label{Fig1}
\end{figure}

{\it Framework.} We will study a physical situation like the one illustrated 
in Fig.\ \ref{Fig1}. A quantum system, described by a Hamiltonian ${\cal H}$, 
is affected by an external static force of strength $\varepsilon$. This force 
gives rise to an additional potential energy, described by an operator $\cal O$ 
\cite{remark}. Due to the presence of a (macroscopically large and weakly 
coupled) heat bath at temperature $T$, thermalization to the density matrix
\begin{equation}
\rho = e^{-\beta ({\cal H} - \varepsilon {\cal O})} /  {\cal Z}
\label{rho_nq} 
\end{equation}
eventually occurs \cite{kubo1991}, where $\beta = 1/T$ ($k_B = 1)$ denotes the 
inverse temperature and ${\cal Z} = \text{Tr} [ e^{-\beta ({\cal H} - 
\varepsilon {\cal O})}]$ is the partition function.
If the perturbation $\varepsilon$ is sufficiently small, the static expectation 
value $\langle {\cal O}(0)\rangle = \text{Tr} [ \rho(0) {\cal O} ]$ is expected 
to be a linear function of $\varepsilon$, i.e., $\langle {\cal O}(0) \rangle 
= \langle {\cal O} \rangle_\text{eq} + \varepsilon \chi(0)$, where $\langle 
\bullet \rangle_\text{eq} = \text{Tr} [ \rho_\text{eq} \bullet ]$ is an 
expectation value with respect to the density matrix $\rho_\text{eq} 
= e^{-\beta {\cal H}}/{\cal Z}_\text{eq}$ and the partition function
${\cal Z}_\text{eq} = \text{Tr} [  e^{-\beta {\cal H}}]$. The static 
susceptibility $\chi(0)$ is given by a Kubo scalar product \cite{kubo1991}
\begin{equation}
\chi = \beta(\Delta {\cal O}; {\cal O}) = \int_0^\beta \text{d} \lambda \,
\text{Tr} [\rho_\text{eq} \, \Delta {\cal O}(-\imath \lambda) {\cal O} 
] 
\end{equation}
with $\Delta {\cal O} = {\cal O} - \langle {\cal O} \rangle_\text{eq}$ 
and $\Delta {\cal O}(-\imath \lambda) = e^{\lambda {\cal H}} \Delta 
{\cal O} \, e^{-\lambda {\cal H}}$. If $\varepsilon$ becomes large enough, this 
linear relationship breaks down. In particular, given an operator 
$\mathcal{O}$ with bounded spectrum, $\lim_{\varepsilon \to \infty} \rho$ is a 
projector on the eigenstates of $\cal O$ with the largest eigenvalue 
${\cal O}_\text{max}$. As a consequence, $\lim_{\varepsilon \to \infty} 
\langle {\cal O}(0) \rangle = {\cal O}_\text{max}$. Hence, a convenient 
quantity is the non-equilibrium parameter
\begin{equation}
\zeta(\varepsilon) = \frac{\langle \Delta {\cal O}(0) \rangle}{{\cal 
O}_\text{max} - \langle {\cal 
O} \rangle_\text{eq}} \, .
\end{equation}
This quantity becomes $\zeta(\varepsilon) = 0$ for $\varepsilon = 0$ and 
$\zeta(\varepsilon) = 1$ for $\varepsilon \to \infty$. While 
$\zeta(\varepsilon)$ 
is a natural measure, it might not always be justified to decide solely on this 
measure if a state is close to or far away from equilibrium \cite{SM}.

Let us now consider a ``sudden quench'', where both the external static 
force and the heat bath are removed at time $t = 0$. Then, at times $t > 0$, 
the density matrix $\rho$ is no equilibrium state of the remaining Hamiltonian 
$\mathcal{H}$ and evolves in time according to the von-Neumann equation, 
$\rho(t) = e^{-i\mathcal{H}t} \rho \, e^{i\mathcal{H}t}$ ($\hbar = 1$). 
While this non-equilibrium scenario is generally different to a 
driven quantum system which unitarily evolves w.r.t.\ a perturbed Hamiltonian,
both setups can, in some cases, be related to each other \cite{brenig1989, SM}.

The central goal of this paper is to investigate the time-dependent 
expectation value  $\langle {\cal O}(t) \rangle = \text{Tr} [ \rho(t) 
{\cal O} ]$ as a function of the non-equilibrium parameter 
$\zeta(\varepsilon)$. In the regime of sufficiently small $\zeta(\varepsilon)$, 
we can certainly expect $\langle {\cal O}(t) \rangle = \varepsilon \, 
\chi(t)$, where $\chi(t)$ denotes the linear-response relaxation function 
$\chi(t) = \beta(\Delta {\cal O}; {\cal O}(t))$ \cite{brenig1989,SM}. However, as the theory 
of linear response is generally restricted to this regime, an intriguing 
questions is: How does the time dependence of $\langle {\cal O}(t) 
\rangle$ change outside this regime? While it is surely challenging 
to provide a general answer, two of us have recently shown \cite{richter2017} 
that the time evolution of $\langle {\cal O}(t) \rangle$ can be become 
completely independent of $\zeta(\varepsilon)$ if the involved operator ${\cal 
O}$ is binary, i.e., if it only has two different eigenvalues. In this paper, 
we unveil that such an independence can also occur for other observables, if 
their matrix structure is in accord with the ETH. This impact of the 
ETH on the route to equilibrium is the main result of our work and demonstrates 
its relevance beyond the mere existence of equilibrium.
\begin{figure}[tb]
\includegraphics[width=0.95\columnwidth]{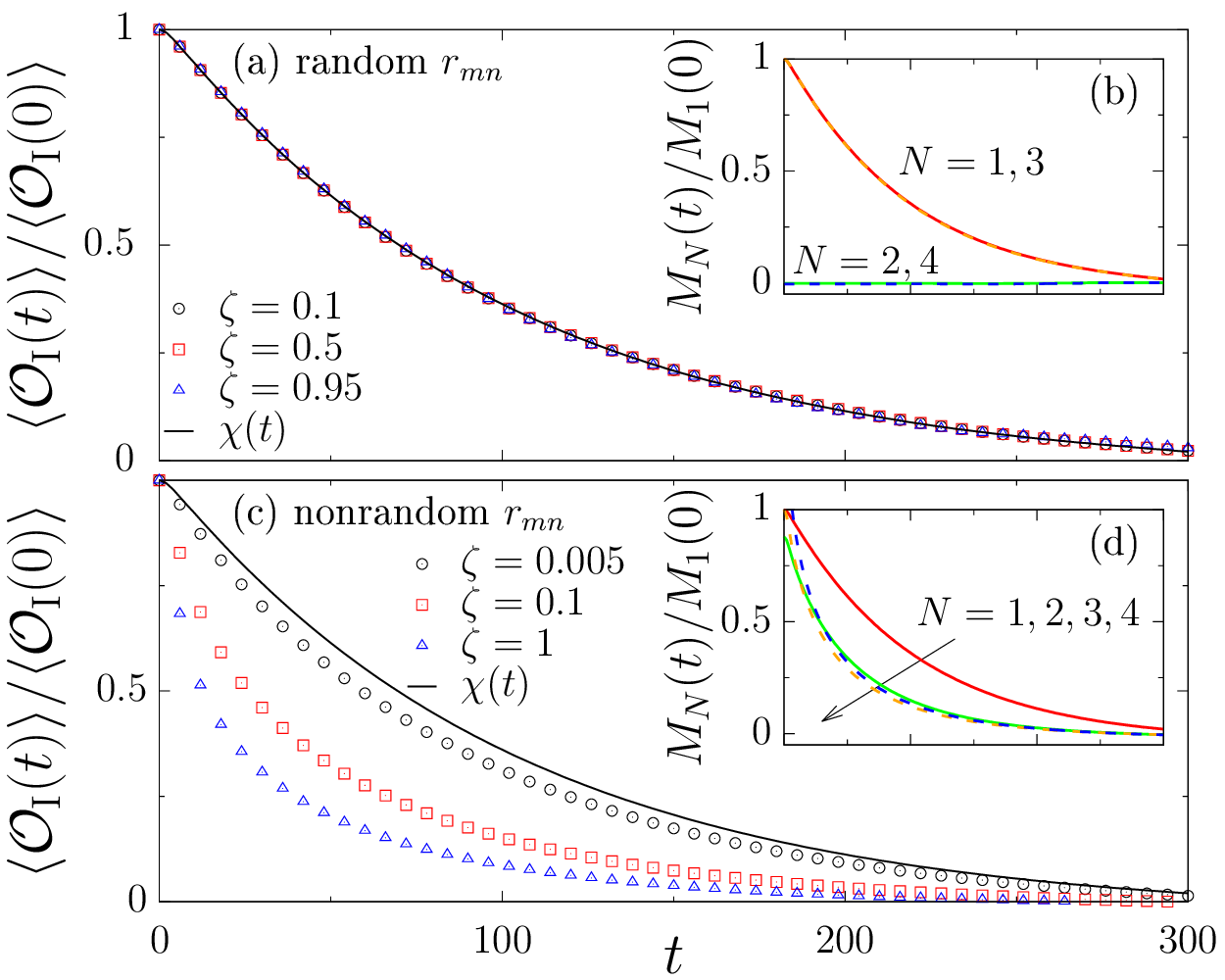}
\caption{(Color online) Numerical simulations for the idealized model. 
Relaxation curve of the normalized expectation value $\langle {\cal 
O}_\text{I}(t) \rangle / \langle {\cal O}_\text{I}(0) \rangle$ for various 
$\zeta(\varepsilon) = \langle {\cal O}_\text{I}(0) \rangle/ {\cal O}_{\text{I}, 
\text{max}}$ as well as the linear-response relaxation function $\chi(t)$. (a) 
random (Gaussian) $r_{mn}$ and (c) non-random (constant) $r_{mn}$. Other  
parameters: $D=1000$, $\Delta E = 1$, $\gamma = \Delta E/100$, $T = 
100$. (b) and (d): $M_N(t) = \text{Tr} [ {\cal O}_\text{I}(t) {\cal 
O}_\text{I}^N ]$ for exponents $N\leq 4$.}
\label{Fig2}
\end{figure}

{\it Illustration: Idealized model.} We begin by illustrating our main result 
for an idealized model of random-matrix type. Its Hamiltonian ${\cal 
H}_\text{I}$ is already given in the diagonal form ${\cal H}_\text{I} = 
\sum_{n=1}^{D} E_n \, | n \rangle \langle n|$, where $D$ is the dimension of the 
Hilbert space and $| n \rangle$ are the eigenstates of ${\cal H}_\text{I}$. The 
corresponding eigenenergies $E_n$ are simply {\it chosen} to be equidistant 
levels $E_n = n \, \Delta E/D$ in a spectrum of width $\Delta E$, i.e., the 
density of states $\Omega = D/\Delta E$ is constant. (For a similar 
model 
with non-constant $\Omega$, see \cite{SM}). In the eigenbasis $| n \rangle$ of 
${\cal H}_\text{I}$, an observable ${\cal O}_\text{I}$ is {\it constructed} as 
${\cal O}_\text{I} = \sum_{m, n =1}^D {\cal O}_{\text{I}, mn} \, | m \rangle \langle n|$ 
with matrix elements
\begin{equation}
{\cal O}_{\text{I}, mn} = \frac{\Gamma}{\sqrt{\gamma^2 + \omega_{mn}^2}} \, 
(1-\delta_{mn}) \, r_{mn} \, , \quad r_{mn} = r_{nm}^* \, , \label{idealized}
\end{equation}
where the frequency $\omega_{mn}$ denotes the energy difference $\omega_{mn} 
= E_n - E_m$ and $\gamma$ is a Lorentzian line width. Thus, for $\gamma 
\ll \Delta E$, ${\cal O}_\text{I}$ is a banded matrix. Moreover, as 
$\delta_{mn}$ is the Kronecker symbol, $\langle {\cal O}_\text{I} 
\rangle_\text{eq} = 0$. Within the constraint of a banded matrix, the complex 
coefficients $r_{mn}$ are free parameters and might be chosen (i) identically
or (ii) randomly, e.g., their real and imaginary part might be drawn 
according to a Gaussian distribution with zero mean. Note that, in 
the latter case (ii), ${\cal O}_\text{I}$ is an ideal realization of the ETH 
ansatz discussed in more detail below. The (in principle irrelevant) prefactor 
$\Gamma$ in Eq.\ \eqref{idealized} is chosen such that $\text{Tr} [ {\cal 
O}_\text{I}^2 ] = \text{Tr} [ {\cal O}_\text{I}^4 ]$.

For the case (ii), we show in Fig.\ \ref{Fig2} (a) the expectation value
$\langle {\cal O}_\text{I}(t) \rangle$, as resulting for the specific initial 
state $\rho$ in Eq.\ (\ref{rho_nq}) and a high temperature $T$. Remarkably, 
when $\langle {\cal O}_\text{I}(t) \rangle$ is normalized to its initial value 
$\langle {\cal O}_\text{I}(0) \rangle$, then the relaxation curve is 
independent of the non-equilibrium parameter $\zeta(\varepsilon) = \langle 
{\cal O}_\text{I}(0) \rangle/ {\cal O}_{\text{I}, \text{max}}$ and coincides 
with the linear-response relaxation function $\chi(t) \propto \text{Tr} [ {\cal 
O}_\text{I}(t) {\cal O}_\text{I}]$ in the entire range of $\zeta(\varepsilon) 
\in 
\, ]0, 1]$ possible. This numerical simulation illustrates our main result: In 
{\it certain} cases, $\chi(t)$ might capture the dynamics at arbitrarily strong 
perturbations. But are such cases generic? This question is non-trivial as 
counterexamples exist. For instance, for the case (i), the relaxation curve in 
Fig.\ \ref{Fig2} (c) depends on $\zeta(\varepsilon)$ and agrees with $\chi(t)$ 
for 
small $\zeta(\varepsilon) \ll 1$ only.

{\it Analytical arguments and ETH.} To work towards an answer to this question, 
let us discuss the initial state $\rho$ in Eq.\ \eqref{rho_nq} for high 
temperatures $T$. For such $T$, we can use the approximation $\rho \approx 
e^{\beta \varepsilon {\cal O}} / \text{Tr} [ e^{\beta \varepsilon \mathcal{O}} 
]$ and a Taylor expansion of this approximation yields
\begin{equation}
\langle {\cal O}(t) \rangle \approx \sum_{N=0}^\infty \alpha_N(\varepsilon) \,
\text{Tr} [{\cal O}(t) {\cal O}^N ] \label{expansion}
\end{equation}
with some $\varepsilon$-dependent coefficients $\alpha_N(\varepsilon)$. Now, 
consider the assumption
\begin{equation}
M_N(t) = \text{Tr} [{\cal O}(t) {\cal O}^N ] \left \{
\begin{array}{ll}
\propto \text{Tr} [{\cal O}(t) {\cal O}] \, , & \text{odd } N \\
= 0 \, , & \text{even } N
\end{array}
\right. \, . \label{assumption}
\end{equation}
If this assumption was justified, the expansion in Eq.\ (\ref{expansion}) 
would directly imply $\langle {\cal O}(t) \rangle \propto \text{Tr} [{\cal O}(t) 
{\cal O}]$. And in fact, as depicted in Fig.\ \ref{Fig2} (b), this assumption 
holds for the idealized model with random $r_{mn}$. In contrast, as shown in 
Fig.\ \ref{Fig2} (d), it breaks down for non-random $r_{mn}$.
Equipped with these prerequisites, we will show that Eq.\ \eqref{assumption}
is closely related to the ETH ansatz \cite{srednicki1994}
\begin{equation}
{\cal O}_{mn} = {\cal F}_\text{d}(\bar{E}) \, \delta_{mn} + 
\Omega(\bar{E})^{-1/2} \, {\cal F}_\text{od}(\bar{E}, \omega_{mn}) \, r_{mn} \ ,
\label{ETH}
\end{equation}
where $\bar{E} = (E_n + E_m)/2$ and ${\cal F}_\text{d}$ and ${\cal 
F}_\text{od}$ are \textit{smooth} functions of their arguments. We will
argue that Eq.\ \eqref{assumption} holds if (qualitatively) ${\cal F}_\text{od}$
varies slowly with $\bar{E}$ and falls of quickly for larger $|\omega_{mn}|$. 
Furthermore, the diagonal elements ${\cal O}_{mm}$ do not need to be smooth 
functions of $\bar{E}$ as claimed by the ETH, variations that are 
uncorrelated with the off-diagonal elements ${\cal O}_{mn}$ leave Eq.\ \eqref{assumption}
valid. We dub this form of ${\cal F}_\text{d}$ and ${\cal F}_\text{od}$ the {\it rigged ETH}. 
As the proof is quite involved for large exponents 
$N$, we restrict ourselves to $N = 2$ and $N =3$ in the following.
For clarity, we also restrict the present consideration to a uniform 
DOS, $\Omega(\bar{E}) = \text{const}.$, and ${\cal F}_\text{od}$ that are independent 
of $\bar{E}$. A full derivation for arbitrary $N$, as well as 
more general $\Omega(\bar{E})$ and ${\cal F}_\text{od}$, 
can be found in the supplemental material \cite{SM}.

We start by writing out the correlation function for $N = 2$ explicitly, 
$\text{Tr} [ {\cal O}(t) {\cal O}^2 ] = \sum_{a,b,c} {\cal O}_{ab} {\cal 
O}_{bc} {\cal O}_{ca} \, e^{\imath \omega_{ab} t}$, and consider a 
Fourier component of this correlation function at fixed frequency $\omega$,
\begin{equation}
\text{Tr} [ {\cal O}(t) {\cal O}^2 ]_\omega = \!\! \sum_{\omega_{ab} = \omega} 
\sum_c {\cal O}_{ab} {\cal O}_{bc} {\cal O}_{ca} \, .
\end{equation}
Given the matrix structure in Eq.\ (\ref{ETH}), the biggest part of the addends 
in the sum are by construction (products of) independent random numbers with 
zero mean.  Thus, to an accuracy set by the law of large numbers, summing the 
latter yields zero as well.  There are, however, index combinations where not 
all factors within the addends have vanishing mean, namely, $c = a$. Focusing on 
these terms, we can write
\begin{equation}
\text{Tr} [ {\cal O}(t) {\cal O}^2 ]_\omega \approx \!\! \sum_{\omega_{ab} = 
\omega} |{\cal O}_{ab}|^2 {\cal O}_{aa} \, . \label{FC2}
\end{equation}
While the numbers $|\mathcal{O}_{ab}|^2$ do not have mean zero, we can, without 
loss of generality, assume that the numbers $\mathcal{O}_{aa}$ have zero mean. 
Because both numbers are independent stochastic variables [cf.\ below Eq.\ \eqref{ETH}], the 
sum in Eq.\ \eqref{FC2} becomes $\text{Tr} [ {\cal O}(t) {\cal O}^2 ]_\omega 
\approx 0$. Since this finding does not depend on $\omega$, we get $\text{Tr} 
[{\cal O}(t) {\cal O}^2 ] \approx 0$, i.e., Eq.\ (\ref{assumption}) for the 
even case $N = 2$.

Now we turn to $N=3$. Here, a Fourier component at fixed frequency $\omega$ 
reads
\begin{equation}
\text{Tr} [ {\cal O}(t) {\cal O}^3 ]_\omega = \!\! \sum_{\omega_{ab} = \omega} 
\sum_{c,d} {\cal O}_{ab} {\cal O}_{bc} {\cal O}_{cd} {\cal O}_{da} \, .
\end{equation}
Again, the contributions of most addends approximately cancel each other upon 
summation.  But again, there are also exceptions, namely, the index combinations 
$c = a$ or $d = b$. Focusing on these terms, we find
\begin{equation}
\text{Tr} [ {\cal O}(t) {\cal O}^3 ]_\omega \approx \!\! \sum_{\omega_{ab} = 
\omega} |{\cal O}_{ab}|^2 \sum_c |{\cal O}_{bc}|^2 + |{\cal O}_{ac}|^2\, . \label{FC4}
\end{equation}
To proceed, recall the matrix structure (Eq.\ \eqref{ETH} and below) and consider the 
above sum over $c$ without the diagonal elements, i.e., $\sum_{c \neq b} |{\cal O}_{bc}|^2+\sum_{c \neq a}|{\cal O}_{ac}|^2$.
While these sums do not vanish, they are practically independent of $a,b$, 
if ${\cal F}_\text{od}(\bar{E}, \omega_{mn})$ is independent of $\bar{E}$ 
and vanishes quickly enough for larger $|\omega_{mn}|$. Thus, the respective 
sums may be replaced by a constant $C$, i.e., $\sum_c |\mathcal{O}_{bc}|^2+|{\cal O}_{ac}|^2 \approx C+|{\cal O}_{aa}|^2 + |{\cal O}_{bb}|^2$. 
The $|{\cal O}_{aa(bb)}|^2$ may be split into their mean and variations: $|{\cal O}_{aa(bb)}|^2:=\bar{{\cal O}^2_\text{d}} + \delta_{a(b)}$. Inserting the above findings into
Eq.\ \eqref{FC4} and exploiting the assumed ``uncorrelatedness'' of the $\delta_{a(b)}$ with the $|{\cal O}_{ab}|^2$ yields 
$\text{Tr}[{\cal O}(t) {\cal O}^3]_\omega \approx (C + 2\bar{{\cal O}^2_\text{d}}) \sum_{\omega_{ab} = 
\omega} |{\cal O}_{ab}|^2$. Comparing this to the exact relation $\text{Tr} [{\cal 
O}(t) {\cal O} ]_\omega = \sum_{\omega_{ab} = \omega} |{\cal O}_{ab}|^2$ and realizing that all findings are independent of $\omega$, eventually yields
$\text{Tr} [{\cal O}(t) {\cal O}^3 ] \propto \text{Tr} [ {\cal O}(t) \mathcal{O}]$, i.e., Eq.\ \eqref{assumption} for the odd case $N = 3$.
Noting that the calculations for $N > 3$ are in principle analogous, we have 
shown that the assumption in Eq.\ (\ref{assumption}) essentially follows from the ansatz in and below Eq.\ \eqref{ETH}.
Hence, we have identified the rigged ETH as the physical mechanism responsible 
for the numerical observation $\langle {\cal O}(t) \rangle \propto \text{Tr} 
[{\cal O}(t) {\cal O}]$, even at strong perturbations.

{\it Illustration: Generic quantum many-body systems.} Let us now illustrate 
the relevance of our results to generic quantum many-body systems. A prototype 
model in this context is the spin-$1/2$ XXZ chain. We hence consider
the Hamiltonian $\mathcal{H}_\text{XXZ} = J \sum_{l=1}^L h_l$ (with periodic 
boundary conditions),
\begin{equation}
h_l = S_l^x S_{l+1}^x + S_l^y S_{l+1}^y + \Delta S_l^z S_{l+1}^z + \Delta' 
S_l^z S_{l+2}^z \, , \label{XXZ}
\end{equation}
where $S_l^{x,y,z}$ are spin-$1/2$ operators at lattice site $l$ and $J = 1$ 
is the antiferromagnetic exchange coupling. For vanishing 
next-nearest-neighbor interaction $\Delta' = 0$, this model is integrable 
in terms of the Bethe ansatz, whereas integrability is broken for any 
$\Delta' \neq 0$. As an observable, we choose the well-known spin current
\cite{heidrichmeisner2007}
\begin{equation}\label{Eq_SCur}
{\cal J} = \Gamma \sum_{l=1}^L S_l^x S_{l+1}^y - S_{l+1}^x S_l^y \, ,
\end{equation}
an important quantity in the context of transport. This quantity we study for 
$\Delta'=0$ and $\Delta = 0.5$, where it is partially conserved 
\cite{ilievski2016, heidrichmeisner2007}, as well as for $\Delta = \Delta' = 
0.5$, where it is expected to fully decay. Generally, $\langle {\cal J} 
\rangle_\text{eq} = 0$, and again, $\text{Tr} [{\cal J}^2] = \text{Tr} 
[{\cal J}^4]$ by a corresponding choice of the prefactor $\Gamma$.

\begin{figure}[tb]
\includegraphics[width = 0.95\columnwidth]{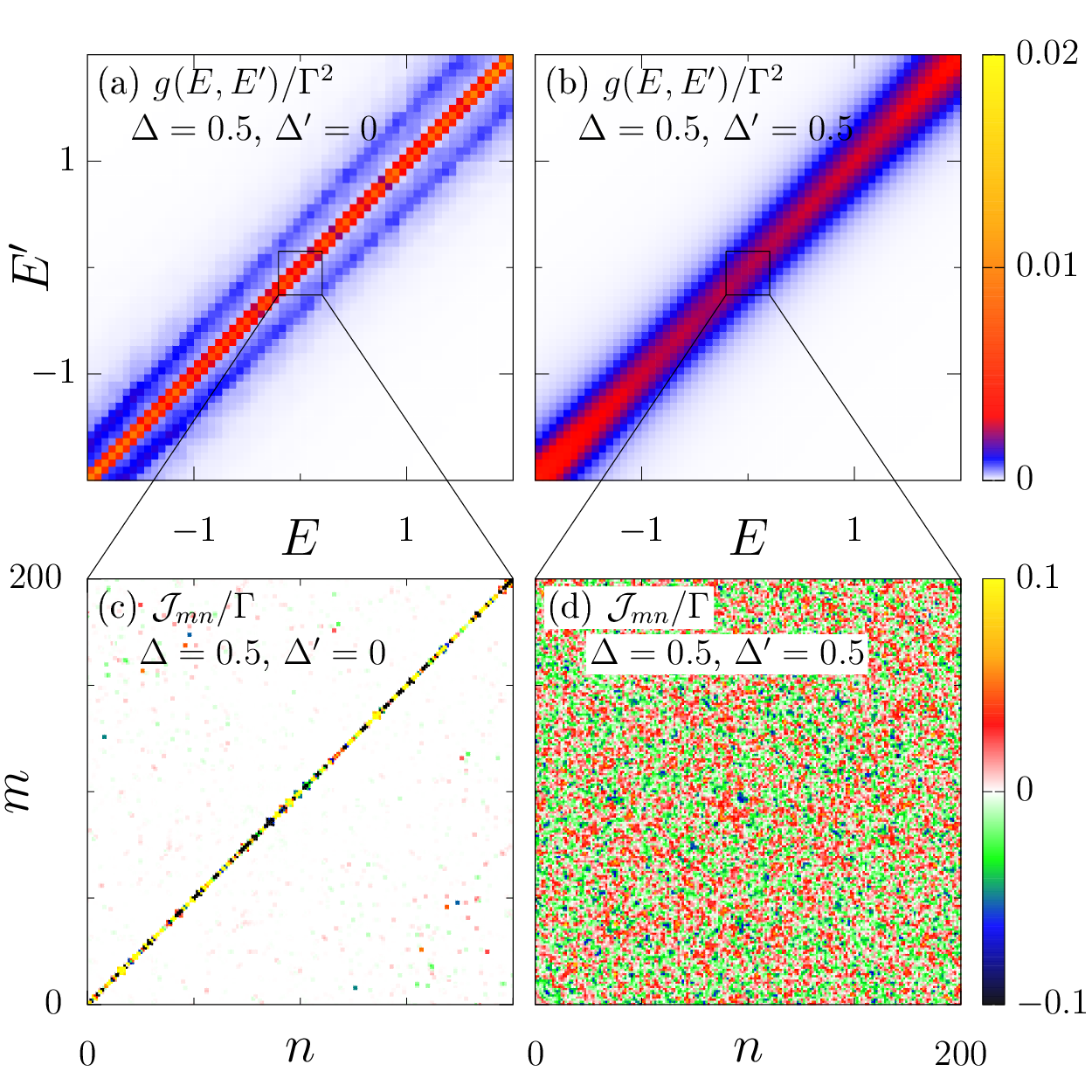}
\caption{(Color online) Matrix elements of $\mathcal{J}$ in the eigenbasis of
${\cal H}_\text{XXZ}$ for (a), (c) integrable case ($\Delta'=0$, $\Delta = 
0.5$) and (b), (d) non-integrable case ($\Delta'=0.5$, $\Delta = 0.5$) in a 
single symmetry subspace ($S^z = 1$ and $k = 1$). Upper row: Coarse grained 
structure $g(E,E')$. Lower row: Close-up of $200 \times 200$ matrix elements 
around the diagonal. In all cases, $L = 20$.}
\label{Fig3}
\end{figure}

The matrix representation of $\mathcal{J}$ in the eigenbasis of 
$\mathcal{H}_\text{XXZ}$ is summarized in Fig.\ \ref{Fig3}. The general 
structure is visualized in Figs.\ \ref{Fig3} (a) and (b) by the use of a 
suitable coarse graining according to
\begin{equation}\label{Eq_gEE}
g(E,E') = \frac{\sum_{mn} |{\cal J}_{mn}|^2 D(\bar{E})}{D(E) D(E^\prime)} \, ,
\end{equation}
where the sum runs over matrix elements ${\cal J}_{nm}$ in two energy 
shells of width $2\delta E$, $E_n \in [E - \delta E,E + \delta E]$ and $E_m \in 
[E^\prime -\delta E,E^\prime+\delta E]$. $D(E)$, $D(E')$, and $D(\bar{E})$ 
denote the number of states in these energy windows. Note that the 
coarse grained quantity $g(E,E')$ can be interpreted as a measure of the 
distribution function ${\cal F}_\text{od}$ in Eq.\ (\ref{ETH}), i.e.,\ 
$g(E,E') \propto |{\cal F}_\text{od}(\bar{E},\omega_{mn})|^2$. Apparently, the 
situation is very similar for the integrable and non-integrable case: Weight 
is concentrated around the diagonal and quickly vanishes further away from the 
diagonal. However, the close-up of matrix elements $\mathcal{J}_{mn}$ in 
Figs.\ \ref{Fig3} (c) and (d) unveils clear differences. On the hand, in the 
integrable case, substantial weight lies directly on the diagonal
and the vast majority of all off-diagonal matrix elements are exactly zero, 
e.g., due to conservation laws. On the other hand, for the non-integrable case, 
the matrix elements appear to be randomly distributed and it is difficult to 
recognize any structure at all \cite{Steinigeweg2013,Beugeling2015}.
\begin{figure}[tb]
\includegraphics[width = 0.95\columnwidth]{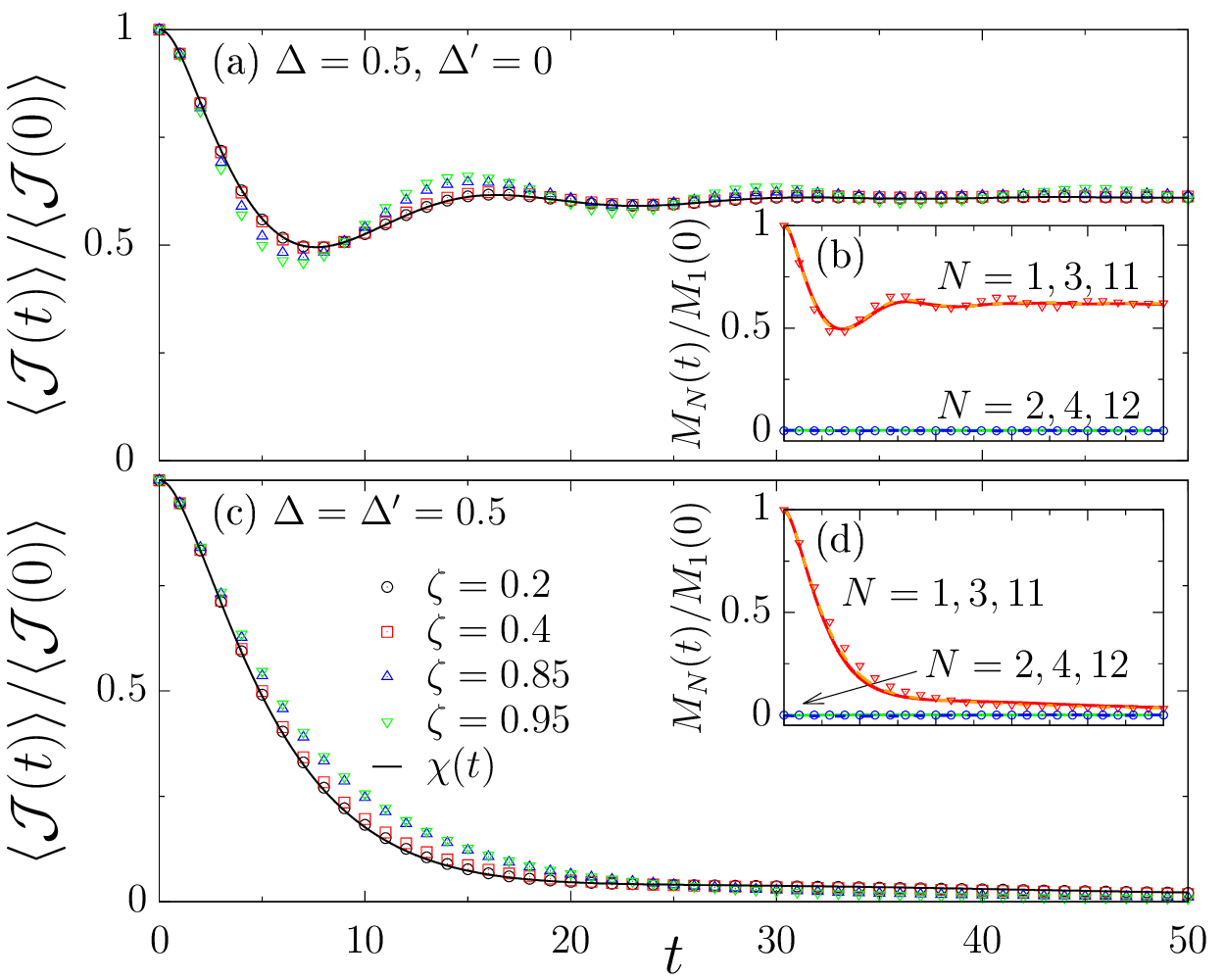}
\caption{(Color online) Numerical simulations for the XXZ spin-$1/2$ chain.
Decay curve of the normalized expectation value $\langle {\cal J}(t) \rangle / 
\langle {\cal J}(0) \rangle$ for various $\zeta(\varepsilon) = \langle {\cal 
J}(0) \rangle/ {\cal J}_\text{max}$ as well as the linear-response relaxation 
function $\chi(t)$. (a) integrable case ($\Delta'=0$, $\Delta = 0.5$) and (c) 
non-integrable case ($\Delta'=0.5$, $\Delta = 0.5$). Other parameters: $L=26$, 
$T = 100$. (b) and (d): $M_N(t) = \text{Tr} [ {\cal J}(t) {\cal J}^N ]$ for 
exponents $N = 1 - 4$ ($L = 20$) and considerably larger $N = 11, 12$ 
\cite{rescaling}.}
\label{Fig4}
\end{figure}

Let us finally turn to the dynamics of ${\cal J}$. (For details on the numerics see 
e.g.\ Refs.\ \cite{sugiura2012,elsayed2013,steinigeweg2014,deraedt2006,weisse2006} and \cite{SM}). 
In Fig.\ \ref{Fig4} we 
depict the expectation value $\langle {\cal J}(t) \rangle$, as resulting for 
the specific initial state $\rho$ in Eq.\ (\ref{rho_nq}) and a high 
temperature $T$. When $\langle {\cal J}(t) \rangle$ is normalized to its initial 
value $\langle {\cal J}(0) \rangle$ again, then the dynamics does not depend
significantly on the non-equilibrium parameter $\zeta(\varepsilon) = 
\langle {\cal J}(0) \rangle/ {\cal J}_\text{max}$ and agrees very well with the 
linear-response relaxation function $\chi(t) \propto \text{Tr} [ {\cal J}(t) 
{\cal J}]$ for all $\zeta(\varepsilon) \in \, ]0, 1]$, as shown in Fig.\ 
\ref{Fig4} (a) for the integrable case  and in Fig.\ \ref{Fig4} (b) for the 
non-integrable case. Although the agreement for large $\zeta(\varepsilon)$ is 
certainly not as perfect as in the idealized model in Fig.\ \ref{Fig2} (a), it 
is still convincing. An explanation for the visible deviations might be 
imperfections in the matrix structure of ${\cal J}$, i.e., small variations 
compared to the rigged ETH ansatz (cf.\ Eq.\ \eqref{ETH} and below). 
Such variations can manifest themselves for instance in correlations 
between the off-diagonal matrix elements, as suggested to exist for local  
operators \cite{Foini2018}.

To substantiate that our analytical arguments apply to this model as well,
we depict $\text{Tr} [ {\cal J}(t) {\cal J}^N ]$ in Figs.\ \ref{Fig4} (b) and 
(d). For all exponents $N$ calculated, we confirm that Eq.\ \eqref{assumption} 
holds to very good accuracy. While the small deviations for large 
$\zeta(\varepsilon)$ in Figs.\ \ref{Fig4} (a) and (c) are presumably caused by 
high orders in $N$, Figs. \ref{Fig4} (b) and 
(d) illustrate that the validity of Eq.\ \eqref{assumption} approximately 
persists even for rather large $N = 11, 12$. (For a more detailed discussion of 
the limitations and physical implications of Eq.\ \eqref{assumption} see 
\cite{SM}). Finally, we should stress that Eq.\ \eqref{assumption} 
is apparently found to apply also in the integrable case, 
even though the diagonal part of the ETH breaks down.

{\it Conclusion.} Given our analytical arguments and our numerical case 
studies, we conclude that the ETH has impact on the route to equilibrium: If 
the off-diagonal part of the ETH applies, then the relaxation process can 
become independent of whether or not the initial state $\rho$ is close to 
equilibrium. Since we have proven this fact for the specific $\rho$ in 
Eq.\ (\ref{rho_nq}) and high temperatures $T$, promising research directions 
include other $\rho$ and lower $T$. Our findings agrees with 
earlier work by Srednicki~\cite{srednicki1999}. 

\textit{Acknowledgements.} This work has been 
funded by the Deutsche Forschungsgemeinschaft (DFG) - Grants No. 397107022 (GE 
1657/3-1), No. 397067869 (STE 2243/3-1), No. 355031190 - within the DFG 
Research Unit FOR 2692. We sincerely thank the members of FOR 2692 for fruitful 
discussions.


\clearpage

\setcounter{figure}{0}
\setcounter{equation}{0}
\renewcommand*{\citenumfont}[1]{S#1}
\renewcommand*{\bibnumfmt}[1]{[S#1]}
\renewcommand{\thefigure}{S\arabic{figure}}
\renewcommand{\theequation}{S\arabic{equation}}

\section{Supplemental Material}

\section{Numerical approach}

In our paper, we use numerical methods to calculate expectation 
values $\langle {\cal O}(t) \rangle$. While we use exact diagonalization 
(ED) to obtain the data for the idealized model in Fig.\ \ref{Fig1}, ED becomes 
costly for the XXZ spin-$1/2$ chain because (i) the Hilbert-space dimension $D 
= 2^L$ grows exponentially with the number of spins $L$ and (ii) the 
calculation of the expectation value $\langle {\cal O}(t) \rangle$ requires ED 
of both, the pre-quench Hamiltonian ${\cal H} - \varepsilon {\cal O}$ and the 
post-quench Hamiltonian ${\cal H}$. Thus, to obtain the data in Figs.\ 4 (a) 
and (c), we proceed differently and employ the concept of dynamical quantum 
typicality as a numerical method \cite{sugiura2012_S, elsayed2013_S, 
steinigeweg2014_S}.
Specifically, we construct a non-equilibrium pure state of 
the form \cite{richter2017_S}
\begin{equation}
\ket{\psi(0)} = \sqrt{\rho} \, \ket{\varphi} / \sqrt{ 
\bra{\varphi}\rho\ket{\varphi} }\ ,
\end{equation}
to mimic the density matrix $\rho$. Here, the reference pure state 
$\ket{\varphi}$ 
is prepared according to the unitary invariant Haar measure, i.e.,
\begin{equation}
\ket{\varphi} = \sum_{k=1}^D c_k \ket{\varphi_k} \, ,
\end{equation}
where the real and imaginary part of the coefficients $c_k$ are both drawn 
from a Gaussian distribution with zero mean and $\ket{\varphi_k}$ denote the 
states 
of our working basis, i.e., the Ising basis. Then, the expectation value 
$\langle {\cal O}(t) \rangle$ can be written as \cite{richter2017_S}
\begin{equation}
\langle {\cal O} (t) \rangle = \bra{\psi(t)} {\cal O} \ket{\psi(t)} + 
f(\ket{\phi}) \label{typicality}
\end{equation}
with the statistical error $f(\ket{\varphi}) \propto 1/\sqrt{D}$ in the limit 
of 
high temperatures $\beta \rightarrow 0$. Thus, $f(\ket{\varphi})$ is negligibly 
small for medium-sized lattice sizes $L$ already.
The main advantage of Eq.\ (\ref{typicality}) stems from the fact, that the 
action of the exponentials $e^{-\beta({\cal H} - \varepsilon {\cal O})}$ 
and $e^{-\imath {\cal H} t}$ can be conveniently evaluated by a forward 
propagation of pure states in imaginary time $\beta$ or real time $t$. For this
forward propagation, various sophisticated methods can be used, e.g., Trotter 
decompositions \cite{deraedt2006_S} or Chebyshev polynomials 
\cite{weisse2006_S}. In the present paper, it is sufficient to apply 
a fourth-order Runge-Kutta scheme \cite{elsayed2013_S, steinigeweg2014_S} with 
a 
small time step $\delta t$. Since this scheme does not require ED and the 
involved operators usually feature a sparse matrix representation, we can 
easily calculate data for $L = 26$ sites, as done in Figs.\ \ref{Fig4} (a) and 
(c).

\section{Relation to other non-equilibrium scenarios}

In the main part of this paper, we have studied the relaxation dynamics 
$\langle {\cal O}(t) \rangle$, as resulting from an initial state $\rho = 
\exp[-\beta({\cal H}-\varepsilon {\cal O})]/ {\cal Z}$. While this setup might 
not be the most common preparation scheme, it can be related to other 
non-equilibrium scenarios in the regime of linear response, i.e., 
small perturbations $\varepsilon$. To this end, let us consider a thermal 
initial state $\rho(-\infty) = \rho_\text{eq}$ and an external field which is 
turned on at $t=-\infty$ and switched off at $t = 0$, i.e.,
\begin{equation}
{\cal H}(t) = \begin{cases}
{\cal H} - \varepsilon(t) \, {\cal O} \, ,  &t < 0 \\
{\cal H} \, , &t\geq 0
\end{cases} \, .
\end{equation}
In this case, the expectation value $\langle {\cal O}(t) \rangle$ can be 
written as 
\begin{equation}
\langle {\cal O}(t) \rangle = \int_{-\infty}^0 \text{d}t' \, \phi(t-t') \,
\varepsilon(t') \, ,
\end{equation} 
where we have introduced the {\it response function} $\phi(t)$. If we now 
assume a weak and quasi-static external field $\varepsilon(t) \approx 
\varepsilon$, 
then this setup translates into the relaxation experiment considered in the 
main part of this paper. Specifically, the relaxation function $\chi(t)$ and 
the response function $\phi(t) = -\beta(\Delta {\cal O}; \dot{{\cal O}}(t))$ 
can be  related according to~\cite{Brenig1989_S}
\begin{equation}
\chi(t) = \Theta(t) \Big [\chi(0) - \int_0^t \text{d}t' \, \phi(t') \Big] \, . 
\end{equation} 
Hence, knowledge of either $\chi(t)$ or $\phi(t)$ is sufficient to describe 
both scenarios. 

\section{Characterization of initial states}

In the main text, we have investigated initial states $\rho = 
\exp[-\beta({\cal H} - \varepsilon {\cal O})] / {\cal Z}$ and argued that the 
tuning of the perturbation $\varepsilon$ allows for a preparation of $\rho$
close to as well as far away from equilibrium. While it seems to be natural
that a $\rho$ with larger $\varepsilon$ has to be considered as further away 
from equilibrium, it is still somewhat ambiguous without reasonable criteria to
characterize $\rho$ as \textit{close to} or \textit{far away from} equilibrium. 
In the following, let us discuss this point in more detail. 

As already mentioned, the initial expectation value $\langle {\cal O}(0) 
\rangle$ can be used as a natural criterion to decide whether a state is far 
away from equilibrium. Recall that $\langle {\cal O}(0) \rangle$ is 
limited by the maximum eigenvalue ${\cal O}_\text{max}$ of the 
operator ${\cal O}$. Therefore, in the main text, we have defined the 
relative deviation from equilibrium
\begin{equation}
\zeta (\varepsilon) = \frac{\langle \Delta {\cal O}(0) \rangle}{{\cal 
O}_\text{max} - \langle {\cal O}_\text{eq} \rangle} \, , 
\end{equation}
with $\langle {\cal O} \rangle_\text{eq} = 0$ in the two case studies. Thus, 
one might call a state $\rho$ close to equilibrium if $\zeta \approx 0$ and far 
from equilibrium if $\zeta \approx 1$.

\begin{figure}[tb]
\centering
\includegraphics[width=\columnwidth]{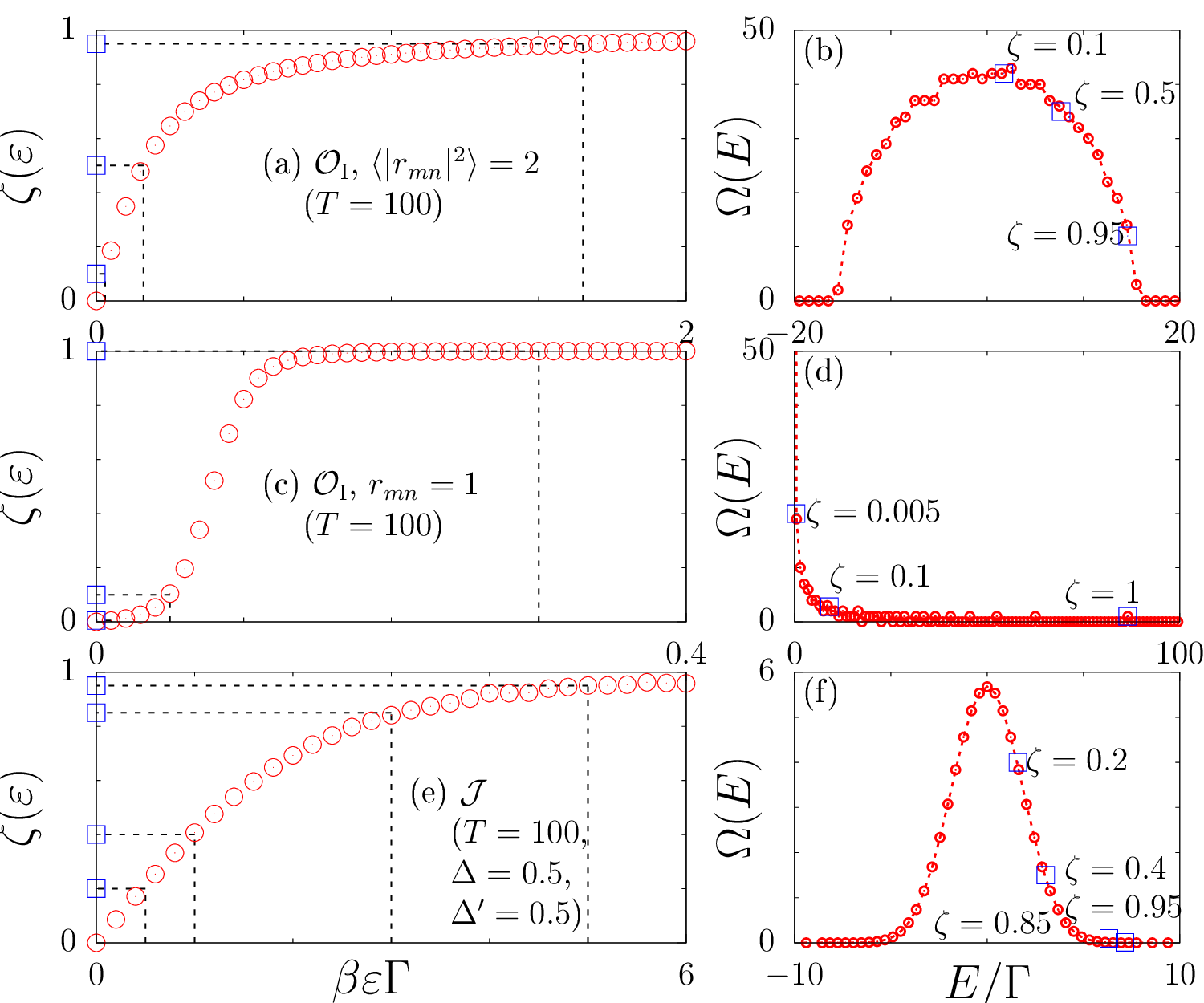}
\caption{(Color online) (a), (c), and (e): Initial expectation value 
$\zeta = \langle {\cal O}(0) \rangle/{\cal O}_\text{max}$ versus perturbation 
$\varepsilon$ for ${\cal O} = {\cal O}_\text{I}$ with random and non-random 
$r_{mn}$ as well as ${\cal O} = {\cal J}$. The lines indicate those values of 
$\varepsilon$ which are used in the main text to study the actual dynamics.
The squares indicate the corresponding values of $\zeta(\varepsilon)$. (b), 
(d), and (f): DOS for the same three operators. The location of
$\zeta(\varepsilon)$ is depicted again by squares.}
\label{FigS3}
\end{figure}

Let us discuss the dependence of $\zeta(\varepsilon)$ on the strength of the 
perturbation $\varepsilon$. In Figs.\ \ref{FigS3} (a), (c), and (e), $\zeta 
(\varepsilon)$ is shown for the idealized operator ${\cal O}_\text{I}$ with
random and non-random $r_{mn}$ as well as for the spin-current operator 
$\mathcal{J}$. In all cases, the temperature is set to $T = 100$. Since 
the static curves at such high $T$ practically do not depend on $\mathcal{H}$, 
it is sufficient to show $\langle {\cal J}(0) \rangle$ only for one choice of 
the anisotropies $\Delta$ and $\Delta'$. For illustration, Figs.\ \ref{FigS3} 
(a), (c), and (e) indicate those values of $\varepsilon$ which are chosen in 
the main text to study the actual dynamics. While we observe that, for 
all observables, $\zeta(\varepsilon)$ monotonically increases with increasing 
$\varepsilon$ until it eventually saturates at $\langle {\cal O}(0) \rangle = 
{\cal O}_\text{max}$, we also find that the values of $\varepsilon$ to 
reach this maximum significantly depend on the specific operator. The latter 
fact becomes clear if one takes into account the different scaling of the 
horizontal $\varepsilon$ axis for the three operators. Apparently, a specific 
value of $\varepsilon$ may cause a large response for one operator but only 
a small response for another operator. Thus, compared to the bare value of 
$\varepsilon$, the parameter $\zeta(\varepsilon)$ yields much better 
information on whether $\rho$ is close to or far away from equilibrium. 
However, in addition to $\zeta(\varepsilon)$, it might be even more insightful 
to consider the whole spectrum and analyze the density of states (DOS) of the 
respective operators.

The DOS of some operator ${\cal O}$ is defined as
\begin{equation}
\Omega(E) = \sum_{n = 1}^D \delta(E - O_n)\ , 
\end{equation}
where the $O_n$ denote the eigenvalues of ${\cal O}$. Even though it is 
straightforward to calculate $\Omega(E)$ by means of exact 
diagonalization, we additionally use a typicality approach \cite{Hams2000_S} to 
evaluate the DOS of ${\cal J}$. Specifically, we exploit
\begin{equation}
\text{Tr}[ e^{-\imath {\cal O} t} ] \approx \bra{\Phi} e^{-\imath {\cal O} t} 
\ket{\Phi}
\end{equation}
for a pure state $\ket{\Phi}$ drawn at random. Using the integral 
representation of the $\delta$-function,
\begin{equation}
\delta(E - O_n) = \frac{1}{2 \pi} \int_{-\infty}^\infty e^{\imath t 
(E - O_n)} \, ,
\end{equation}
we then have
\begin{equation}
\Omega(E) \approx C \int_{-\Theta}^\Theta \text{d}t \, e^{\imath O t} 
\bra{\Phi} e^{-\imath {\cal O} t} \ket{\Phi} \ ,
\label{DOSTyp}
\end{equation}
where $C$ is a normalization constant and the spectral resolution $\delta E$ 
depends on the cutoff time, $\delta E = \pi/\Theta$. In Figs.\ \ref{FigS3} (b), 
(d) and (f) we depict the DOS for ${\cal O}_\text{I}$ with random and 
non-random $r_{mn}$ and for ${\cal J}$. Also in the case of $\Omega(E)$, we 
observe clear differences between the three operators. First, for ${\cal 
O}_\text{I}$ with random $r_{mn}$, $\Omega(E)$ follows the well-known 
semi-circle from random matrix theory and second, for ${\cal O}_\text{I}$ with 
non-random $r_{mn}$, $\Omega(E)$ exhibits a strong degeneracy around $E = 0$ 
and a long tail up to a quite large maximum eigenvalue $\mathcal{O}_\text{max}$.
Third, in case of ${\cal J}$, $\Omega(E)$ has a Gaussian shape.

As an orientation, we indicate also in Figs.\ \ref{FigS3} (b), (d) and (f) the 
location of the initial states from the text with their different values of 
$\zeta(\varepsilon)$. While this location is roughly at the maximum of 
$\Omega(E)$ for small $\zeta(\varepsilon)$, it shifts to the borders of the 
spectrum as $\zeta(\varepsilon)$ is increased to larger values. In particular, 
for 
$\zeta(\varepsilon) \approx 1$, we find that the initial states are located in 
a region with a very low DOS and it is certainly justified to consider such 
states as far from equilibrium. Hence, our choices of $\zeta(\varepsilon)$ 
cover the whole range of initial states close to and far away from equilibrium.

\section{Random-matrix model with non-constant density of states}

In the context of Fig.\ \ref{Fig2}, we considered a random-matrix model 
in order to demonstrate the validty of our analytical results for an ideal 
realization 
of the ETH Ansatz given in and below Eq.\ \eqref{ETH}. In this case, the 
Hamiltonian ${\cal H}_\text{I}$ was chosen with equidistant energy levels, 
i.e., a constant density of states $\Omega = \text{const}.$ Here, we 
additionally study the non-equilibrium dynamics 
of a similar random-matrix model with non-constant $\Omega$. Specifically, the 
eigenenergies $E_n$ of ${\cal H}_\text{I}$ are now randomly 
drawn from an exponential probability distribution in the interval $E_n \in 
[0,1]$. Thus, the density of states in this interval is 
approximately given by
\begin{equation}
 \Omega(E) \propto e^{\beta' E}\ .
\end{equation}
We here choose $\beta' = 1$ such that $\Omega(1) \approx e \cdot 
\Omega(0)$, and $\Omega(E)$ significantly deviates from a constant. 
The respective observable ${\cal O}_\text{I}$ is again constructed as described 
in Eq.\ \eqref{idealized}, with 
random or non-random elements 
$r_{mn}$ and Lorentzian damping. 
However, due to the non-constant density of states, 
the matrix elements with mean energy $\bar{E}$ are now additionally multiplied 
by the factor 
$\Omega(\bar{E})^{-1/2} = e^{-\beta' \bar{E}/2}$, cf.\ Eq.\ \eqref{ETH}. 

\begin{figure}[tb]
 \centering
 \includegraphics[width=0.95\columnwidth]{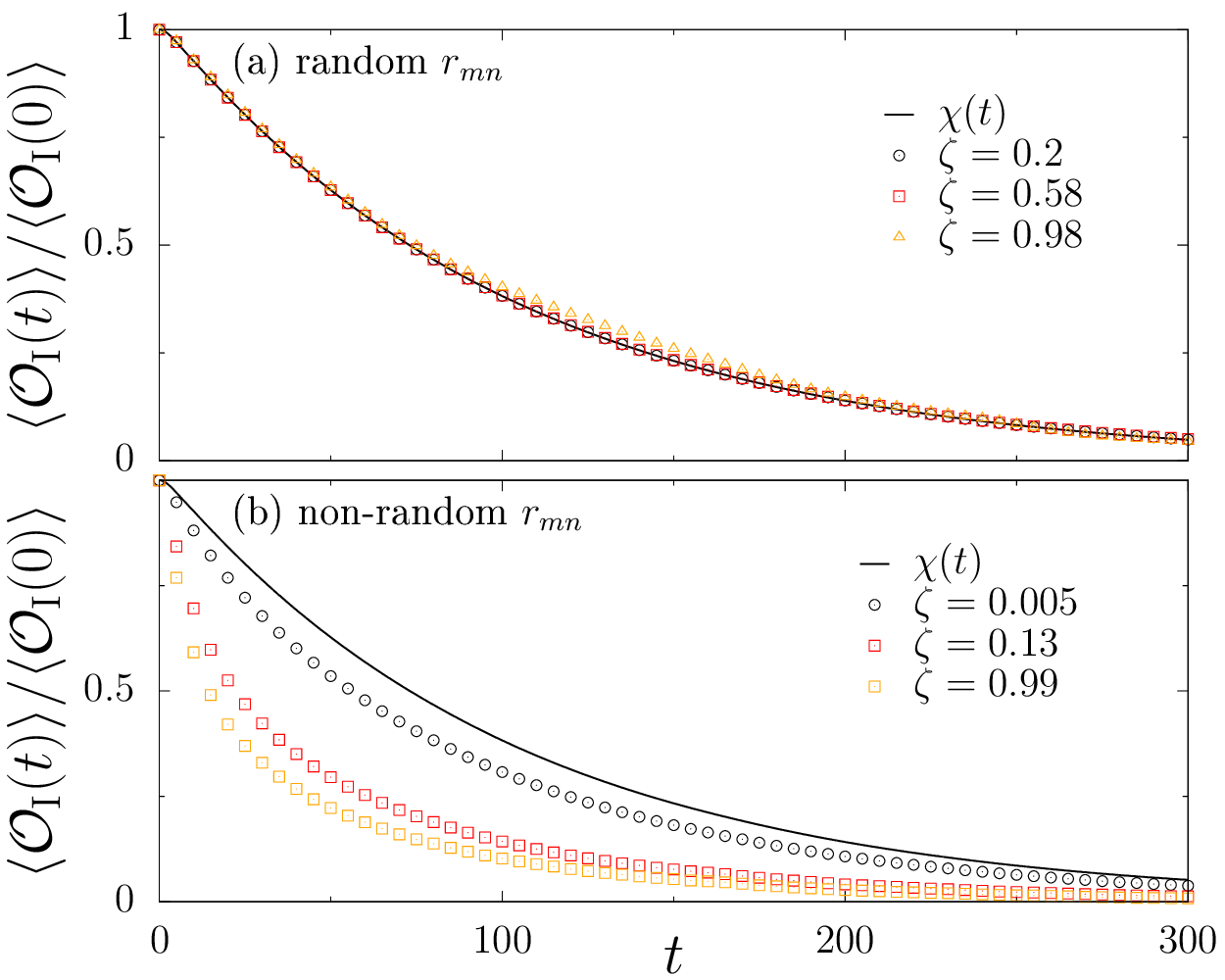}
 \caption{(Color online) Numerical simulations for the random-matrix model with 
exponential 
 density of states. Relaxation curve of the normalized expectation value 
$\langle {\cal 
O}_\text{I}(t) \rangle / \langle {\cal O}_\text{I}(0) \rangle$ for various 
$\zeta(\varepsilon) = \langle {\cal O}_\text{I}(0) \rangle/ {\cal O}_{\text{I}, 
\text{max}}$ as well as the linear-response relaxation function $\chi(t)$. (a) 
random (Gaussian) $r_{mn}$ and (b) non-random (constant) $r_{mn}$. Other  
parameters: $D=2000$, $\Delta E = 1$, $\gamma = \Delta E/100$, $T = 
100$.}
 \label{FigS2}
\end{figure}

In Fig.\ \ref{FigS2} we depict the non-equilibrium dynamics for the 
random-matrix model with non-constant density of states. 
Generally, the results are very similar compared to the previous model studied 
in the context of Fig.\ \ref{Fig2}.
In particular, for random matrix elements $r_{mn}$, the non-equilibrium dynamics 
$\langle {\cal O}_\text{I}(t)\rangle$ is
essentially proportional to the linear response relaxation function in the 
entire regime $\zeta \in ]0,1]$. This behavior 
clearly breaks down if the $r_{mn}$ are non-random, cf.\ Fig.\ \ref{FigS2} (b). 
While there are some small deviations in Fig.\ \ref{FigS2} (a)
for the very large $\zeta = 0.98$, let us emphasize that our derivations
generally rely on the law of large numbers,
and, therefore, the results in Fig.\ \ref{FigS2} are still rather convincing in 
view of the small matrix dimension $D = 2000$. 
We elaborate in detail on the reasons for (and the limits of) the good agreement 
in Fig.\ \ref{FigS2} (a) in the last section of 
the supplemental material. However, important features for the occurrence of 
this agreement are a density of states that is well 
described by an exponential over a wide energy range and a decay of $\langle 
{\cal O}(t)\rangle$ 
that is rather slow on a timescale set by the inverse of the above wide energy 
range.

\section{Implications of the main assumption on physically relevant quantities}

While the validity of Eq.\ \eqref{assumption} from the main text is clearly 
mathematically nontrivial for any $N>1$, it is not evident whether or not a
validity for, say, $N=5$ entails a physically relevant result. Specifically, if 
the number of particles (spins) $N_p$ increases, 
does the minimum $N$ in Eq.\ \eqref{assumption}, which is required to yield a 
physically meaningful result, increase as well? And if so how? 
Before elaborating on this question we anticipate a brief answer:
Let the spectrum of the observable ${\cal O}$ be Gaussian, with a standard 
deviation $\sigma_{\cal O}$. 
Then, the validity of Eq.\ \eqref{assumption} up to a maximum exponent $N=f$ 
renders
\begin{equation}
\label{statement}
\langle {\cal O}(t)\rangle \propto \text{Tr}[{\cal O}(t){\cal O}(0)]\  
\end{equation}
valid, for initial states 
\begin{equation}
 \label{highte}
 \rho \propto \exp(\beta \varepsilon {\cal O})\ , 
\end{equation}
and initial values $\langle {\cal O}(0)\rangle$ which are on the order of $f 
\sigma_{\cal O}$. Hence, even if one finds that $f$ does not scale with $N_p$ 
at 
all, but assumes a value like, e.g., $N=5$, this already implies the 
validity of Eq.\ (\ref{statement}) for the largest part of the spectrum. It 
implies furthermore a validity of Eq.\ \eqref{statement} which is, loosely 
speaking, five times larger 
than what linear response alone accounts for. However, an $f$ that does not 
scale with $N_p$ (and is not very large ($\gg 1$) already for small systems) 
does not entail a 
validity of Eq.\ \eqref{statement} in the ``large-deviation regime'', i.e., in 
the outer tails of the spectrum of ${\cal O}$. 
Note that a principal consideration on the expected $f$ for a given 
observable and Hamiltonian is provided in the next section.

We now embark on a detailed justification of the above statement.
In essence, Eq.\ \eqref{statement} will be valid if the following truncation of 
the Taylor-series approximates 
the respective exponential well (we set $\alpha = \beta \varepsilon$ in the 
following),
\begin{equation}
\label{trunc}
 e^{\alpha {\cal O}} \approx \sum_n^{n=f} \frac{1}{n !}(\alpha {\cal O})^n\ .
\end{equation}
The question now is: how large may $\alpha, {\cal O}$ maximally be for a given 
$f$
to render Eq.\ \eqref{trunc} valid? The weight in the addends of a 
Taylor-series is Poisson distributed. If $\alpha {\cal O}$ was a number, 
the expectation value and standard deviation of the respective 
``term distribution'' would accordingly (see standard math textbooks) read
\begin{equation}
\label{poisson}
 \bar{n} = \alpha {\cal O}, \quad \sigma_n =\sqrt{ \alpha {\cal O}}\ .
\end{equation}
Thus at 
\begin{equation}
\label{limit1}
\alpha {\cal O} \leq \frac{f}{c_1}\ ,
\end{equation}
with $c_1$ a bit larger than unity, Eq.\ \eqref{trunc} will be valid 
to good accuracy for $f \gg 1$. For small $f$, choosing $c_1 =10$ will suffice.
However, $\alpha {\cal O}$ is not a number. Thus we aim at 
making Eq.\ \eqref{trunc} valid for most of the spectrum of ${\cal O}$. Let 
$E_{\cal O}$ be a (positive) number such that a big percentage of the 
eigenvalues of ${\cal O}$ falls into
$[-E_{\cal O}, E_{\cal O}]$. Then, a mathematically admissible version of 
Eq.\ \eqref{limit1} reads
\begin{equation}
\label{limit2}
\alpha E_{\cal O} \leq \frac{f}{c_1}\ .
\end{equation}
Let us now think of ${\cal O}$ as a standard ``local'' observable. The widths 
of 
the spectra of such observables usually scale as $\propto \sqrt{N_p}$ and the 
spectra are more or less Gaussian. Thus $E_{\cal O}$ may be chosen as $E_{\cal 
O}=c_2 \sqrt{N_p}$, with $c_2$ sufficiently large, see below. 
This makes Eq.\ \eqref{limit2} read $\alpha c_2 \sqrt{N_p} \leq \frac{f}{c_1}$, 
i.e., the regime of 
$\alpha$ for which Eq.\ \eqref{trunc} holds is limited as
\begin{equation}
\label{limit3}
\alpha \leq \frac{f}{c_1 c_2 \sqrt{N_p}}\ .
\end{equation}

Now we analyze the largest initial expectation value, $\langle{\cal 
O}(0)\rangle$, that may be reached based on the above $\alpha$, i.e., a given 
$f$.
Let ${\cal O}$ have a for simplicity a continuous spectrum $s(x)$. As argued 
below Eq.\ \eqref{limit2}, $s$ is expected to be approximately Gaussian with a 
standard deviation 
which may be written as $\sigma_{\cal O} = c_3\sqrt{N_p}$. Hence we get 
$s(x)\propto \exp(-\frac{x^2}{2 c_3^2 N_p})$. Now we come back to the above 
choice of 
$c_2$. If we require Eq.\ \eqref{trunc} to hold for, e.g., $99.73 \%$ of all 
eigenstates of ${\cal O}$, then we have to choose $c_2 = 3 \cdot c_3$. Let 
$p(x)$ be the probability
to find the observable ${\cal O}$ initially at $x$. The most ``dislocated'' 
$p(x)$ we can faithfully generate with some given $f$ in an initial state of 
the form \eqref{highte} is:
\begin{equation}
\label{prob}
p(x)\propto e^{-\frac{x^2}{2 c_3^2 N_p}} e^{\frac{f}{c_1 c_2 \sqrt{N_p}}}\propto 
 e^{-\frac{1}{2 c_3^2 N_p}  (x- \frac{f c_3^2\sqrt{N_p}}{c_1 c_2})^2}\ .  
\end{equation}
The mean value (and the maximum) of this distribution is
\begin{equation}
\label{max}
 \langle{\cal O}(0)\rangle =   \frac{f c_3^2\sqrt{N_p}}{c_1 c_2}\ .  
\end{equation}
Thus, $\langle{\cal O}(0)\rangle$ scaled 
against $\sigma_{\cal O}$ gives
\begin{equation}
\label{end}
 \frac{\langle{\cal O}(0)\rangle}{\sigma_{\cal O}}    =   \frac{f c_3}{c_1 c_2}\ 
,  
\end{equation}
or, imposing the above relation of $c_2, c_3$,
\begin{equation}
\label{end}
 \frac{\langle{\cal O}(0)\rangle}{\sigma_{\cal O}}    =   \frac{f }{3 \cdot c_1 
}\ .  
\end{equation}
Now comes the crucial point: This relation is entirely independent of 
${N_p}$. Hence the validity of Eq.\ \eqref{assumption} entails a physically 
meaningful result even if 
the maximum $N$ for which it holds does not scale with ${N_p}$ at all. Very 
roughly speaking (setting $c_1$ to unity), this means that the validity of 
Eq.\ \eqref{assumption} up to some $N=f$ ensures that Eq.\ \eqref{statement} 
will hold up to initial values of ${\cal O}$ that are $f/3$-times larger 
than the standard deviation of ${\cal O}$, regardless of the size of the system. 
Note that a scaling like, e.g., $f \propto \sqrt{N_p}$ would 
already address the ``large-deviation regime'', i.e., a validity of 
(\ref{statement}) for initial values of ${\cal O}$ that are far out in the 
tails 
of the spectrum of ${\cal O}$.

\section{$\mathrm{Tr}[{\cal O}(t) {\cal O}^N]$ for arbitrary $N$}

To begin with, consider a Fourier component of $\text{Tr}[{\cal O}(t){\cal 
O}^N]$ 
at fixed frequency $\omega$:
\begin{equation} \label{fullsum}
 \text{Tr}[{\cal O}(t){\cal O}^N]_\omega = \hspace{-0.2cm}\sum_{\omega_{a b} = 
\omega} 
 \hspace{-0.1cm}{\cal O}_{a b} \cdot \hspace{-0.3cm}\sum_{i,j,k,l,...}
 \hspace{-0.1cm}( {\cal O}_{b i} {\cal O}_{ij} \cdots  {\cal O}_{kl} {\cal O}_{l 
a})\ ,
\end{equation}
where the second sum is just an expanded representation of $({\cal O}^N)_{b a}$
and the first sum runs over all indices $a$, $b$ with $E_b-E_a = \omega_{ab}$.
Since ${\cal O}$ is essentially a random matrix [see Eq.\ \eqref{ETH}], most 
addends 
in Eq.\ \eqref{fullsum} are products of independent random numbers. As such they 
will 
be random numbers themselves, with random phases (or signs, in case ${\cal O}$ 
should be real). 
Hence, to an accuracy set by the law of large numbers, these addends will sum up 
to zero. 
However, there are index combinations for which the respective addends are not 
just 
products of independent random numbers  but necessarily real and positive. 
(These are also the 
only addends that would ``survive'' an averaging of Eq.\ \eqref{fullsum} over 
concrete implementations of ${\cal O}$.) For the remainder we focus exclusively 
on these addends. 

A first necessary, but by no means sufficient condition on the indices to 
render 
the respective addends surely positive, may be stated as follows: The indices 
of one of the ${\cal O}$'s in the second sum on the r.h.s.\ of Eq.\ 
\eqref{fullsum} 
must be chosen as ${\cal O}_{b a}$. A second  necessary but not sufficient 
condition on 
``surely positive'' addends is that the remaining products of ${\cal O}$'s must 
consist of 
even numbers of factors (elements of ${\cal O}$). This means that the addends of 
the 
second sum on the r.h.s.\ of Eq.\ \eqref{fullsum} must be of the following form
\begin{equation}\label{defpos}
 \underbrace{{\cal O}_{b i} \cdots {\cal O}_{j b}}_{N-P-1} {\cal O}_{b a} 
 \underbrace{{\cal O}_{a k} \cdots {\cal O}_{l a}}_{P}\ ,  
\end{equation}
where the expressions below the underbraces indicate the respective numbers of 
multiplied
elements of ${\cal O}$, $N-P-1$ and $P$ are even integers. Since $N-P-1$ and $P$ 
both must be even,
it follows that $N$ must be odd. This already establishes the lower line of Eq.\ 
\eqref{assumption}. 
However, without any further condition on the remaining indices $i,j,k,l,$ etc., 
the products given 
in \eqref{defpos} are not yet surely positive. A third condition which (together 
with the 
previous two conditions) is sufficient for sure positiveness of the addends in 
Eq.\ \eqref{fullsum} 
is that the remaining indices of the underbraced products must all be 
``paired''. This means that 
if such a product has some factor ${\cal O}_{ij}$ it must {\it also} have the 
factor ${\cal O}_{ji}$ 
such as to form $|{\cal O}_{ji}|^2$. (This principle also underlies the first 
and second necessary condition.)
Such a pairing can be achieved in many ways. For example, for the first product 
in \eqref{defpos} 
some of them may be schematically written as
\begin{eqnarray} \label{pairings}
 &{\cal O}_{b i}{\cal O}_{ij} {\cal O}_{jk} \cdots  {\cal O}_{kj} {\cal O}_{ji} 
{\cal O}_{ i b}\ , \label{pairings1} \\
 &{\cal O}_{b i}{\cal O}_{ib} {\cal O}_{b j} {\cal O}_{ji} \cdots {\cal O}_{ij} 
{\cal O}_{ j b}\ , \label{pairings2}\\
 &{\cal O}_{b i}{\cal O}_{ij}  \cdots  {\cal O}_{ji} {\cal O}_{ i b}{\cal O}_{b 
l} {\cal O}_{ l b}\ . \label{pairings3}
 \end{eqnarray}
The ``building blocks'' of these paired index combinations are sequences of 
indices 
with a mirror symmetry, such that the second part repeats the first part in 
reversed order, 
i.e., $b, i,j,k, \cdots ,k,j,i, b$. The first example, i.e.,\ 
\eqref{pairings1}, 
consists only of one single building block. The second and the third example, 
i.e.,\ \eqref{pairings2} and \eqref{pairings3}, consist of two building 
blocks of different lengths in different order [\eqref{pairings2}: short first; 
\eqref{pairings3}: long first]. 
For large $N-P-1$ and respectively $P$, there are very many possibilities to 
create 
different building blocks of different lengths and arrange them in different 
orders. 
It is very hard to organize these possibilities in a reasonable manner. 
Fortunately, 
we do not need to do this here. To proceed further, we first make an assumption 
on a property of the building blocks and show that  
the validity of Eq.\ \eqref{assumption} may be inferred from this assumption. 
Then we justify the assumption, 
thereby elucidating its limitations. In order to clearly formulate the 
assumption we consider a 
``summed building block'', $f(c,Q)$, given as
\begin{equation}\label{sumblock}
 f(c,Q):=  \sum_{i,j,k,l,...}{\cal O}_{c i}{\cal O}_{ij} {\cal O}_{jk} \cdots  
{\cal O}_{kj} {\cal O}_{ji} {\cal O}_{i c}\ ,
\end{equation}
where $Q$ (integer, even) indicates the numbers of ${\cal O}$'s in the addends. 
Note that the summation 
is over all indices except for the ``end index'', in this case $c$.  Now, the 
assumption is as follows: 
Assume that the $f(c,Q)$ are actually independent of $c$, regardless of $Q$. To 
this end, consider a sum over the terms in \eqref{defpos} as
\begin{align}\label{gsum}
 g(&a,b,N,P) := \nonumber\\ &\left(\sum_{i,...,j}\underbrace{{\cal O}_{b i} 
\cdots {\cal O}_{j b}}_{N-P-1}\right) {\cal O}_{b a} 
 \left(\sum_{k,...,l}\underbrace{{\cal O}_{a k} \cdots {\cal O}_{l 
a}}_{P}\right)\ .  
\end{align}
Note that all relevant contributions of the second sum in Eq.\ \eqref{fullsum}, 
i.e.,\ to  $({\cal O}^N)_{b a}$, 
are of the form $g(a,b, N, P)$. Since the paired contributions to the two sums 
of $g(a,b, N, P)$ all consist of 
summed building blocks of the form $f(c,Q)$, it follows that, given the above 
assumption on the $f(c,Q)$, 
the two sums themselves do not depend on $a,b$ respectively. Hence $g(a,b, N, 
P)$ is of the approximate 
(up to unpaired contributions) form 
\begin{equation}\label{gapr}
 g(a,b, N, P)  \approx C(N,P) \cdot {\cal O}_{b a}\ ,  
\end{equation}
where $C(N,P)$ is a constant w.r.t.\ $a,b$.
Since, as mentioned below Eq.\ \eqref{gsum}, $({\cal O}^N)_{b a}$ is a sum of 
$g(a,b, N, P)$'s over the 
allowed $P$, it follows that  
\begin{equation}\label{propto}
  ({\cal O}^N)_{b a} \propto {\cal O}_{b a} + \text{non-surely pos. 
contribution}\ .
\end{equation}
If Eq.\ \eqref{propto} holds, then Eq.\ \eqref{assumption} from the main text 
also holds, in the 
``law of large numbers sense'' described below Eq.\ \eqref{fullsum}.

Now, the remaining crucial question is if and to what extend $f(c,Q)$ is indeed 
independent of $c$. In order to analyze this, consider a version of the ETH 
ansatz as suggested, e.g., 
in \cite{Srednicki1999_S} and given by Eq.\ \eqref{ETH} with some additional 
conditions on the functions 
$\Omega(E),\ {\cal F}_\text{od}(\bar{E}, \omega)$ contained in that ansatz 
(``rigged ETH'').
Let $\Omega(E)$ be a function that is piecewise nicely described by 
exponentials, i.e, $\Omega(E) \approx \Omega_0 \exp (\beta E)$. 
Let the energy intervals to which the respective mono-exponential form applies 
be not too small fractions 
of the full width of the energy spectrum, such as, e.g., for a Gaussian DOS. Let 
furthermore 
${\cal F}_\text{od}(\bar{E}, \omega)$ be approximately independent of $\bar{E}$ 
within such energy 
intervals, i.e., ${\cal F}_\text{od}(\bar{E}, \omega) \approx {\cal 
F}_\text{od}(\omega)$ for all $\bar{E}$ 
from an interval. Eventually ${\cal F}_\text{od}(\omega)$ must be suitably 
narrow: 
Let $\delta \omega$ be the typical width of ${\cal F}_\text{od}(\omega)$. Then 
we require 
$Q \cdot \delta \omega$ to be still smaller than the interval on which the DOS 
is 
mono-exponential. We dub these specifications of Eq.\ \eqref{ETH} the 
\textit{rigged ETH}. 
This eventually sets a limit to the maximum power of ${\cal O}$, i.e., $N$. As 
one last brutal simplification 
we set ${\cal F}_\text{d}(\bar{E})=0$. This condition can be relaxed 
substantially as we will 
demonstrate in a forthcoming publication. Here, we employ it for simplicity and 
clarity of presentation.

Equipped with these specifications, we now embark on a concrete calculation of 
$f(c,Q)$, 
\begin{equation} \label{diag1}
 f(c,Q) = \sum_{i,j, \cdots, k,l}|{\cal O}_{c i}|^2 |{\cal O}_{ij}|^2 \cdots 
|{\cal O}_{kl}|^2\ .
\end{equation}
We plug Eq.\ \eqref{ETH} together with the aforementioned specifications 
into Eq.\ \eqref{diag1}. Furthermore, we entirely rely on the law of large 
numbers, i.e., we replace 
$|r_{ij}|^2 \rightarrow 1$. This yields
\begin{align} \label{diag2}
 f&(c,Q) \approx\sum_{i,j, \cdots, k,l}  \Omega_0^{-Q} 
 e^{-\frac{\beta}{2}( E_c + 2 E_i + \cdots + 2 E_k + E_l)} \\
 &\cdot {\cal F}_\text{od}^2( E_c -E_i)  {\cal F}_\text{od}^2( E_i -E_j) \cdots 
{\cal F}_\text{od}^2( E_k -E_l)\ . \nonumber
\end{align}
Now, we go from sums to integrals, essentially by plugging in the respective 
DOS's
\begin{align} \label{diag3}
 f&(c,Q) \approx \int \Omega_0^{-Q} e^{-\frac{\beta}{2}( E_c + 2 E_1 + \cdots + 
2 E_{L-1} + E_L)} \\
 &\cdot {\cal F}_\text{od}^2( E_c -E_1)  {\cal F}_\text{od}^2(E_1 -E_2) \cdots 
{\cal F}_\text{od}^2(E_{Q-1} - E_Q) \nonumber \\
 &\cdot \Omega_0^{Q} e^{\beta (E_1 + E_2 +\cdots + E_{Q-1} + E_Q)} {\text d}E_1 
{\text d}E_2 \cdots {\text d}E_Q\ . \nonumber
\end{align}
Here, we off course heavily rely on the above specifications. A closer look 
reveals 
that most of the DOS's from the ETH ansatz cancel nicely with the DOS's from the 
integrations:
\begin{align} \label{diag4}
 f&(c,Q) \approx e^{-\frac{\beta}{2}E_c} \int e^{\frac{\beta}{2} E_Q} \\
 &\cdot  {\cal F}_\text{od}^2( E_c -E_1)  {\cal F}_\text{od}^2( E_1 -E_2) \cdots 
{\cal F}_\text{od}^2( E_{Q-1} - E_Q) \nonumber \\
 &\cdot {\text d}E_1 {\text d}E_2  \cdots {\text d}E_Q\ . \nonumber
\end{align}
We apply the following  linear change of variables,  
\begin{eqnarray} \label{transform}
 & E_1 , E_2 \cdots E_{Q-1}, E_Q& \\
 &\rightarrow& \nonumber \\
 &\omega_1 :=  E_c - E_1, \,\omega_2 :=  E_1 -E_2,  \cdots   & \nonumber \\
 &\omega_Q :=  E_{Q-1} - E_Q \nonumber\ , 
\end{eqnarray}
which features a Jacobian 
determinant which equals unity. (The matrix has lower triangle form and the 
diagonal
elements are all $-1$.) Furthermore, we realize
\begin{equation} \label{aha}
 E_Q = E_c -\sum_{i=1}^Q \omega_i\ .
\end{equation}
Thus, in the new variables the integral from Eq.\ \eqref{diag4} reads
\begin{eqnarray} \label{diag5}
 f(c,Q) &\approx  &  \int  e^{\frac{-\beta}{2} \sum_{i=1}^Q \omega_i   } \\
 &\cdot& {\cal F}_\text{od}^2(\omega_1)  {\cal F}_\text{od}^2(\omega_2) \cdots 
{\cal F}_\text{od}^2(\omega_Q) \nonumber \\
 &\cdot& {\text d} \omega_1 {\text d} \omega_2 \cdots  {\text d} \omega_Q   
\nonumber  \\
 &=& \left(\int e^{\frac{-\beta \omega}{2}} {\cal F}_\text{od}^2(\omega){\text 
d} \omega \right)^Q\ . \nonumber 
\end{eqnarray}
Displayed in this form, $f(c,Q)$ is manifestly independent of $c$. 
This completes the justification for the assumption made before Eq.\ 
\eqref{gsum}.

Thus, to sum up, the validity of Eq.\ \eqref{assumption} from the main text can 
be expected, 
more or less pronounced, if the following specifications, in addition to the 
standard ETH ansatz, apply:
(i) DOS's that are in accord with exponentials over substantial energy ranges, 
(ii)  ${\cal F}_\text{od}(\bar{E}, \omega)$'s that vary slowly or/and weakly 
with $\bar{E}$, 
(iii) ${\cal F}_\text{od}(\bar{E}, \omega)$'s that are narrow in $\omega$ 
direction, 
(iv) small $N$. This appears to be in accord with the numerical findings: For 
the idealized 
random-matrix model, (i)-(iii) apply very well. Thus, the above arguments apply 
even for large $N$ 
or exponential initial states featuring large exponents. For the lattice 
spin-model, (i)-(iii) apply also,
but not as strictly as for the random-matrix model. Thus, Eq.\ 
\eqref{assumption} from the main text
breaks down for exponential initial states featuring large exponents.


\begin{thebibliography}{99}

\bibitem{polkovnikov2011}
A. Polkovnikov, K. Sengupta, A. Silva, and M. Vengalattore, 
Rev. Mod. Phys. {\bf 83}, 863 (2011).

\bibitem{eisert2015}
J. Eisert, M. Friesdorf, and C. Gogolin,
Nat. Phys. {\bf 11}, 124 (2015). 

\bibitem{nandkishore2015}
R. Nandkishore and D. A. Huse,
Annu. Rev. Condens. Matter Phys. {\bf 6}, 15 (2015).

\bibitem{gogolin2016}
C. Gogolin and J. Eisert,
Rep. Prog. Phys. {\bf 79}, 056001 (2016).

\bibitem{dalessio2016}
L. D'Alessio, Y. Kafri, A. Polkovnikov, and M. Rigol,
Adv. Phys. {\bf 65}, 239 (2016).

\bibitem{bloch2005}
I. Bloch, 
Nat. Phys. {\bf 1}, 23 (2005).

\bibitem{langen2015}
T. Langen, R. Geiger, and J. Schmiedmayer, 
Ann. Rev. Condens. Matter Phys. {\bf 6}, 201 (2015).

\bibitem{schollwoeck2011}
U. Schollw\"ock,
Rev. Mod. Phys. {\bf 77}, 259 (2005);
Ann. Phys. {\bf 326}, 96 (2011).

\bibitem{popescu2006}
S. Popescu, A. J. Short, and A. Winter, 
Nat. Phys. {\bf 2}, 754 (2006).

\bibitem{goldstein2006}
S. Goldstein, J. L. Lebowitz, R. Tumulka, and N. Zangh\`i,
Phys. Rev. Lett. {\bf 96}, 050403 (2006).

\bibitem{reimann2007}
P. Reimann, 
Phys. Rev. Lett. {\bf 99}, 160404 (2007).

\bibitem{deutsch1991}
J. M. Deutsch, 
Phys. Rev. A {\bf 43}, 2046 (1991). 

\bibitem{srednicki1994}
M. Srednicki, 
Phys. Rev. E {\bf 50}, 888 (1994).

\bibitem{rigol2005}
M. Rigol, V. Dunjko, and M. Olshanii, 
Nature {\bf 452}, 854 (2008).

\bibitem{khatami2013}
E. Khatami, G. Pupillo, M. Srednicki, and M. Rigol, 
Phys. Rev. Lett. {\bf 111}, 050403 (2013). 

\bibitem{reimann2016}
P. Reimann, 
Nat. Commun. {\bf 7}, 10821 (2016).

\bibitem{garciapintos2017}
L. P. Garc\'{\i}a-Pintos, N. Linden, A. S. L. Malabarba, A. J. Short, and A. 
Winter,
Phys. Rev. X {\bf 7}, 031027 (2017).

\bibitem{kubo1991}
R. Kubo, M. Toda and, N. Hashitsume,
{\it Statistical Physics II: Nonequilibrium Statistical Mechanics}, 
Solid-State Sciences {\bf 31} (Springer, Berlin, 1991).

\bibitem{srednicki1999}
M. Srednicki, 
J. Phys. A {\bf 32}, 1163 (1999).

\bibitem{remark}
Not for every observable ${\cal O}$ there is a force. See, e.g., 
R. Zwanzig, Annu. Rev. Phys. Chem. {\bf 16}, 67 (1965).

\bibitem{brenig1989}
W. Brenig, 
{\it Statistical Theory of Heat: Nonequilibrium Phenomena} (Springer, Berlin, 
1989).

\bibitem{richter2017}
J. Richter and R. Steinigeweg, 
Phys. Rev. E {\bf 99}, 012114 (2019).

\bibitem{heidrichmeisner2007}
F. Heidrich-Meisner, A. Honecker, and W. Brenig, 
Eur. Phys. J. Spec. Top. \textbf{151}, 135 (2007).

\bibitem{ilievski2016}
E. Ilievski, M. Medenjak, T. Prosen, and L. Zadnik,
J. Stat. Mech. {\bf 2016}, 064008 (2016).

\bibitem{Steinigeweg2013}
R. Steinigeweg, J. Herbrych, and P. Prelov\v{s}ek, 
Phys. Rev. E {\bf 87}, 012118 (2013).

\bibitem{Beugeling2015}
W. Beugeling, R. Moessner, and M. Haque, 
Phys. Rev. E \textbf{91}, 012144 (2015).

\bibitem{sugiura2012}
S. Sugiura and A. Shimizu, 
Phys. Rev. Lett. \textbf{108}, 240401 (2012).

\bibitem{elsayed2013}
T. A. Elsayed and B. V. Fine, 
Phys. Rev. Lett. {\bf 110}, 070404 (2013). 

\bibitem{steinigeweg2014}
R. Steinigeweg, J. Gemmer, and W. Brenig,
Phys. Rev. Lett. {\bf 112}, 120601 (2014).

\bibitem{deraedt2006}
H. De Raedt and K. Michielsen,
{\it Handbook of Theoretical and Computational Nanotechnology} (American 
Scientific Publishers, Los Angeles, 2006).

\bibitem{weisse2006}
A. Wei\ss e, G. Wellein, A. Alvermann, and H. Fehske, 
Rev. Mod. Phys. {\bf 78}, 275 (2006).

\bibitem{Foini2018}
L. Foini and J. Kurchan, 
Phys. Rev. E {\bf 99}, 042139 (2019).

\bibitem{SM}
See supplemental material for details. 

\bibitem{rescaling}
Note that the data for $N = 11$ has been multiplied by a factor to ensure 
$M_N(0)/M_1(0) = 1$ for easier comparison.

\end{thebibliography}

\begin{thebibliography}{99}

\bibitem{sugiura2012_S}
S. Sugiura and A. Shimizu, 
Phys. Rev. Lett. \textbf{108}, 240401 (2012).

\bibitem{elsayed2013_S}
T. A. Elsayed and B. V. Fine, 
Phys. Rev. Lett. {\bf 110}, 070404 (2013).

\bibitem{steinigeweg2014_S}
R. Steinigeweg, J. Gemmer, and W. Brenig,
Phys. Rev. Lett. {\bf 112}, 120601 (2014).

\bibitem{richter2017_S}
J. Richter and R. Steinigeweg, 
Phys. Rev. E {\bf 99}, 012114 (2019).

\bibitem{deraedt2006_S}
H. De Raedt and K. Michielsen,
{\it Handbook of Theoretical and Computational Nanotechnology} (American 
Scientific Publishers, Los Angeles, 2006).

\bibitem{weisse2006_S}
A. Wei\ss e, G. Wellein, A. Alvermann, and H. Fehske, 
Rev. Mod. Phys. {\bf 78}, 275 (2006).

\bibitem{Brenig1989_S}
W. Brenig, 
{\it Statistical Theory of Heat: Nonequilibrium Phenomena} (Springer, Berlin, 
1989).

\bibitem{Hams2000_S}
A. Hams and H. De Raedt, 
Phys. Rev. E {\bf 62}, 4365 (2000).

\bibitem{Srednicki1999_S}
M. Srednicki, 
J. Phys. A {\bf 32}, 1163 (1999).

\end{thebibliography}
 \end{document}